\documentclass[twocolumn,showpacs,prl,floatfix]{revtex4}
\usepackage{graphicx}
\usepackage{amsmath}
\def\beq{\begin{eqnarray}} \def\eeq{\end{eqnarray}}
\def\beqstar{\begin{eqnarray*}} \def\eeqstar{\end{eqnarray*}}
\def\beg{\begin{gather}}
\newcommand{\bal}{\begin{align}} \def\eal{\end{align}}
\newcommand{\beqe}{\begin{equation}} \newcommand{\eeqe}{\end{equation}}
\newcommand{\p}[1]{(\ref{#1})}
\newcommand{\tbf}{\textbf}

\newcommand{\tcal}{\cal}

\begin {document}
\title{Two-body correlation functions
\protect\\in   nuclear matter with $np$ condensate}
\author{ A. A. Isayev}
 \affiliation{Kharkov Institute of
Physics and Technology, Academicheskaya Str. 1,
 Kharkov, 61108, Ukraine
 \\
 Kharkov National University, Maidan Svobody, 4, Kharkov, 61077, Ukraine
 }
 \date{\today}
\begin{abstract}   The density, spin and isospin correlation functions in
nuclear matter with a neutron-proton ($np$) condensate are
calculated to study the possible signatures of the BEC-BCS
crossover in the low-density region. It is shown that the
criterion of the crossover (Phys. Rev. Lett. {\bf 95}, 090402
(2005)), consisting in the change of the sign of the density
correlation function at low momentum transfer, fails to describe
correctly the density-driven BEC-BCS transition at finite isospin
asymmetry or finite temperature. As  an unambiguous  signature of
the BEC-BCS transition, there can be used the presence (BCS
regime) or absence (BEC regime) of the singularity in the momentum
distribution of the quasiparticle density of states.
\end{abstract}
 \maketitle


\maketitle

\tbf {Introduction}. Neutron-proton pairing correlations play an
important role in a number of contexts~\cite{KBT,BB}, including
the study of medium mass $N\approx Z$ nuclei produced at the
radioactive nuclear beam facilities~\cite{G} and the process of
deuteron formation in  medium-energy heavy ion
collisions~\cite{BLS}.
 For not too
low densities, $np$ pairing correlations crucially depend on the
overlap between neutron and proton Fermi surfaces and even small
isospin asymmetry effectively destroys a condensate with $np$
Cooper pairs due to the Pauli blocking effect~\cite{AFRS,SL,AIPY}.
However, under decreasing density, when neutrons and protons start
to bind in deuterons and the spatial separation between deuterons,
and between deuterons and extra neutrons is large, the Pauli
blocking loses its efficiency in destroying a $np$ condensate. In
such a situation, despite the fact that the isospin asymmetry may
be very large, a $np$ condensate survives and exists in the form
of a Bose-Einstein condensate of deuterons.

The transition from BCS superconductivity to Bose--Einstein
condensation (BEC) occurs in a Fermi system, if either density is
decreased or the attractive interaction between fermions is
increased sufficiently. This transition was studied, first, in
superconducting semiconductors~\cite{E} and then in an attractive
Fermi gas~\cite{NS}. Later it was realized that an analogous phase
transition takes place in symmetric nuclear matter, when $np$
Cooper pairs at higher densities go over to Bose--Einstein
condensate of deuterons at lower densities~\cite{BLS,AFRS}. During
this transition the chemical potential changes its sign at certain
critical density (Mott transition), approaching  half of the
deuteron binding energy at ultra low densities. In
Ref.~\cite{AFRS} crossover from $np$ superfluidity to BEC of
deuterons was investigated in the $T$--matrix approach, where the
pole in the $T$--matrix determines the critical temperature of BEC
of bound states in the case of negative chemical potential and the
critical temperature of the appearance of $np$ Cooper pairs in the
case of positive chemical potential. The influence  of isospin
asymmetry on  the BEC-BCS crossover in  nuclear matter was studied
in Ref.~\cite{LNS} within the BCS formalism. It has been shown
that Bose-Einstein condensate is weakly affected by an additional
gas of free neutrons even at very large asymmetries. The same
conclusion was also confirmed  in Ref.~\cite{IYB} on the base of
the variational approach for the thermodynamic potential.

The recent upsurge of interest to the BEC-BCS crossover is caused
by finding the BCS pairing in ultracold trapped quantum atom
gases~\cite{ZSS,KHG}. In this study we examine the possible
signatures of the BEC-BCS crossover in low-density nuclear matter.
It may have interesting consequences, for example, in  the far
tails of the density profiles of exotic nuclei, where a deuteron
condensate can exist in spite of the fact that  the density there
can be quite asymmetric. Besides, similar physical effects can
play an important role in expanding nuclear matter, formed in
heavy ion collisions, or in nuclear matter in the crust of a
neutron star. The main emphasis is laid on the behavior of the
density, spin and isospin correlation functions across the BEC-BCS
transition region. The study is motivated by the results of
Ref.~\cite{MGB}, where the authors state that the density
correlation function of a two-component ultracold fermionic gas of
atoms
 changes sign at low momentum transfer and this represents
an unambiguous signature of the BEC-BCS crossover. This statement
is checked for nuclear matter taking into account  additional
factors: finite isospin asymmetry or finite temperature. In both
cases, this criterion fails to provide a correct description of
the density-driven BEC-BCS crossover and cannot serve as the
universal feature of transition between two states of the system.

\tbf{Basic equations.} Superfluid states of nuclear matter are
described
  by the normal $f$ and anomalous $g$ distribution functions of
  nucleons
  \beqe f_{\kappa_1\kappa_2}=\mbox{Tr}\,\varrho
  a^+_{\kappa_2}a_{\kappa_1},\; g_{\kappa_1\kappa_2}=\mbox{Tr}\,\varrho
a_{\kappa_2}a_{\kappa_1},\label{2}\end{equation} where
$\kappa\equiv({\bf{k}},\sigma,\tau)$, $\bf k$ is momentum,
$\sigma(\tau)$ is the projection of spin (isospin) on the third
axis, $\varrho$ is the density matrix of the system.
 We shall
study $np$ pairing correlations in the pairing channel with
 total spin $S$ and
isospin $T$ of a pair $S=1$, $T=0$ and  the projections
$S_z=T_z=0$.
 In this case the distribution functions for isospin asymmetric
 nuclear matter have the structure
 \bal f({\bf k})&= f_{00}({\bf k})\sigma_0\tau_0+f_{03}({\bf
k})\sigma_0\tau_3
,\label{7.0}\\
g({\bf k})&=g_{30} ({\bf k})\sigma_3\sigma_2\tau_2,\nonumber
\end{align}
where $\sigma_i$ and $\tau_k$ are the Pauli matrices in spin and
isospin spaces, respectively. Using the minimum principle of the
thermodynamic potential and procedure of block
diagonalization~\cite{AIPY}, one can obtain expressions for the
distribution functions \bal f_{00}({\bf
k})&=\frac{1}{2}-\frac{\xi_k}{4E_k}\Bigl(\tanh\frac{E_k^+}{2T}
+\tanh\frac{E_k^-}{2T}\Bigr),  \label{6'}\\
f_{03}({\bf k})&=\frac{1}{4}\Bigl(\tanh\frac{E_k^+}{2T}
-\tanh\frac{E_k^-}{2T}\Bigr),\label{6}\\
g_{30}({\bf k})&=-\frac{\Delta({\bf
k})}{4E_k}\Bigl(\tanh\frac{E_k^+}{2T}
+\tanh\frac{E_k^-}{2T}\Bigr).\label{6''}\end{align} Here \beq
E_k^\pm&=E_k\pm\delta\mu=\sqrt{\xi^2_k+\Delta^2({\bf
k})}\pm\delta\mu,\; \xi_k=\frac{{\bf k}^2}{2m}-\mu,
\label{6'''}\end{eqnarray} $\Delta$ being the energy gap in the
quasiparticle excitation spectrum, $m$ being the effective nucleon
mass,   $\mu$ and $\delta\mu$ being half of a sum and half of a
difference of neutron and proton chemical potentials,
respectively.

   Equations,
governing $np$ pairing correlations in the $S=1,\,T=0$ pairing
channel, can be obtained on the base of Green's function formalism
and have the form~\cite{SL,AIPY,IYB}
  \beg \Delta({\bf k})
=-\frac{1}{V}\sum_{{\bf k}'}V({\bf k},{\bf k}')\frac{\Delta({\bf
k}')}{2E_{k'}}(1-f(E_{k'}^+)-f(E_{k'}^-)), \label{8}\\
\varrho=\frac{2}{V}\sum_{\bf
k}\Bigl(1-\frac{\xi_k}{E_k}[1-f(E_k^+)-f(E_k^-)]\Bigr)\equiv\frac{2}{V}\sum_{\bf
k}n_k,\label{10}\\
\alpha\varrho=\frac{2}{V}\sum_{\bf
k}\Bigl(f(E_k^-)-f(E_k^+)\Bigr),\label{11}\end{gather} where
$f(E)$ is  Fermi distribution function.  Eq.~\p{8} is equation for
the energy gap $\Delta$ and Eqs.~\p{10}, \p{11} are equations for
the total density $\varrho=\varrho_p+\varrho_n$ and neutron excess
$\delta\varrho=\varrho_n-\varrho_p\equiv\alpha\varrho$ ($\alpha$
being the asymmetry parameter). Note that, since we consider
unitary superfluid state ($\Delta\Delta^+\propto I$),
Eqs.~\p{8}-\p{11} formally coincide with the equations for a
two-component isospin asymmetric superfluid with singlet spin
pairing between unlike fermions. Introducing the anomalous density
$$\psi({\bf k})=<a^+_{n,k}a^+_{p,-k}>=\frac{\Delta({\bf
k})}{2E_k}\Bigl(1-f(E_k^+)-f(E_k^-)\Bigr)$$ and using Eq.~\p{10},
one can represent Eq.~\p{8} for the energy gap in the form
\beqe\frac{k^2}{m}\psi({\bf k})+(1-n_k)\sum_{{\bf k}'}V({\bf
k},{\bf k}')\psi({\bf k}')=2\mu\psi({\bf
k}).\label{12}\end{equation} In the limit of vanishing density,
$n_k\rightarrow0$, Eq.~\p{12} goes over into the Schr\"odinger
equation for the deuteron bound state~\cite{BLS,LNS}. The
corresponding energy eigenvalue is equal to $2\mu$. The change in
the sign of the mean chemical potential $\mu$
 of neutrons and protons under decreasing density of nuclear matter signals the
transition from the regime of large overlapping $np$ Cooper pairs
to the regime of non-overlapping bound states (deuterons).

Let us consider the two-body density correlation function \bal
{\tcal D} (\tbf {x},\tbf {x}')&=\mbox{Tr}\varrho\Delta\hat
n(\tbf{x})\Delta\hat n(\tbf
{x}'),\; \Delta\hat n(\tbf{x})=\hat n(\tbf{x})-\hat n,\label{12n}\\
\hat
n(\tbf{x})&\equiv\sum_{\sigma\tau}\psi^+_{\sigma\tau}(\tbf{x})\psi_{\sigma\tau}(\tbf{x})
\nonumber
\\ \quad & =
\frac{1}{V}\sum_{\sigma\tau\tbf{kk}'}e^{i(\tbf{k}'-\tbf{k})\tbf{x}}a^+_{\tbf{k}\sigma\tau}
a_{\tbf{k}'\sigma\tau},\;\nonumber\\
\hat n&=
\frac{1}{V}\sum_{\sigma\tau\tbf{k}}a^+_{\tbf{k}\sigma\tau}
a_{\tbf{k}\sigma\tau}.\nonumber \end{align} Its general structure
in the spatially uniform and isotropic case reads~\cite{LL5} \beq
{\tcal D}(\tbf {x},\tbf {x}')=\varrho\delta(\tbf{r})+\varrho
D(r),\; \tbf{r}=\tbf {x}-\tbf {x}'\end{eqnarray} The function
$D(r)$ is called the density correlation function as well. We will
be just interested in the behavior of the function $D(r)$. The
trace in Eq.~\p{12} can be calculated, using definitions \p{2} and
Wick rules. Taking into account Eqs.~\p{7.0} and going to the
Fourier representation
$$D(q)=\int d^3\tbf {r}e^{i\tbf {q}\tbf
{r}}D(r),$$ one can get \beq D(q)=I_g^{30}(q)-
I_f^{00}(q)-I_f^{03}(q),\label{12.3}\end{eqnarray} where \beq
I_f^{00}(q)&=\frac{4}{\pi^3\varrho}\int_0^\infty
dr\,r^2j_0(rq)\Bigl[\int_0^\infty
dk\,k^2f_{00}(k)j_0(rk)\Bigr]^2,\nonumber\\
I_f^{03}(q)&=\frac{4}{\pi^3\varrho}\int_0^\infty
dr\,r^2j_0(rq)\Bigl[\int_0^\infty
dk\,k^2f_{03}(k)j_0(rk)\Bigr]^2,\nonumber\\
I_g^{30}(q)&=\frac{4}{\pi^3\varrho}\int_0^\infty
dr\,r^2j_0(rq)\Bigl[\int_0^\infty
dk\,k^2g_{30}(k)j_0(rk)\Bigr]^2.\nonumber
\end{eqnarray}
Here $j_0$ is the spherical Bessel function of the first kind and
zeroth order. The functions $I_f^{00},I_f^{03}$ and $I_g^{30}$
represent the normal and anomalous contributions to the density
correlation function. Analogously, we can consider the two-body
spin correlation function \bal{{\tcal S}}_{\mu\nu} (\tbf {x},\tbf
{x}')&=\mbox{Tr}\varrho\Delta\hat s_\mu(\tbf{x})\Delta\hat
s_\nu(\tbf
{x}'),\; \Delta\hat s_\mu(\tbf{x})=\hat s_\mu(\tbf{x})-\hat s_\mu,\label{13}\\
\hat s_\mu(\tbf{x})&\equiv\frac{1}{2}\sum_{\sigma\sigma '
\tau}\psi^+_{\sigma\tau}(\tbf{x})(\sigma_\mu)_{\sigma\sigma'}\psi_{\sigma'\tau}(\tbf{x})
\nonumber\\
\quad &=\frac{1}{2V}\sum_{\sigma\sigma ' \tau \tbf{kk}'}
e^{i(\tbf{k}'-\tbf{k})\tbf{x}}a^+_{\tbf{k}\sigma\tau}(\sigma_\mu)_{\sigma\sigma'}
a_{\tbf{k}'\sigma'\tau},\,\nonumber\\
\hat s_\mu&=
\frac{1}{2V}\sum_{\sigma\sigma'\tau\tbf{k}}a^+_{\tbf{k}\sigma\tau}(\sigma_\mu)_{\sigma\sigma'}
a_{\tbf{k}\sigma'\tau},\nonumber
\end{align}
and the two-body isospin correlation function \bal{\tcal
T}_{\mu\nu} (\tbf {x},\tbf {x}')&=\mbox{Tr}\varrho\Delta\hat
t_\mu(\tbf{x})\Delta\hat t_\nu(\tbf
{x}'),\; \Delta\hat t_\mu(\tbf{x})=\hat t_\mu(\tbf{x})-\hat t_\mu,\label{14}\\
\hat t_\mu(\tbf{x})&\equiv\frac{1}{2}\sum_{\sigma\tau
\tau'}\psi^+_{\sigma\tau}(\tbf{x})(\tau_\mu)_{\tau\tau'}\psi_{\sigma\tau'}(\tbf{x})
\nonumber\\
\quad &=\frac{1}{2V}\sum_{\sigma\tau \tau' \tbf{kk}'}
e^{i(\tbf{k}'-\tbf{k})\tbf{x}}a^+_{\tbf{k}\sigma\tau}(\tau_\mu)_{\tau\tau'}
a_{\tbf{k}'\sigma\tau'},\,\nonumber\\
\hat t_\mu&=
\frac{1}{2V}\sum_{\sigma\tau\tau'\tbf{k}}a^+_{\tbf{k}\sigma\tau}(\tau_\mu)_{\tau\tau'}
a_{\tbf{k}\sigma\tau'}.\nonumber
\end{align}
Their general structure for  isospin asymmetric nuclear matter
without spin polarization is \bal {\cal
S}_{\mu\nu}(\tbf{x,x}')&=\frac{\varrho}{4}\,\delta_{\mu\nu}\,\delta(\tbf{r})+\varrho\,
S_{\mu\nu}(r),\label{15}\\
 {\cal
T}_{\mu\nu}(\tbf{x,x}')&=\frac{\varrho}{4}\,\delta_{\mu\nu}\,\delta(\tbf{r})+
\frac{\alpha\varrho}{4}\,i\epsilon_{\mu\nu 3}\,\delta
(\tbf{r})+\varrho\, T_{\mu\nu}(r).\end{align} Then, calculating
traces in Eqs.~\p{13}, \p{14}, for the Fourier transforms of the
spin and isospin correlation functions one can get   \bal
S_{\mu\nu}(q)&=
-\frac{1}{4}\left\{\delta_{\mu\nu}(I_f^{00}(q)+I_f^{03}(q))\right. \label{16}\\
&\quad+ \left.
(\delta_{\mu\nu}-2\delta_{3\mu}\delta_{3\nu})I_g^{30}(q)\right\},\nonumber\\
T_{\mu\nu}(q)&=-
\frac{1}{4}\{\delta_{\mu\nu}(I_f^{00}(q)+I_g^{30}(q))\label{17}\\
&\quad-
(\delta_{\mu\nu}-2\delta_{3\mu}\delta_{3\nu})I_f^{03}(q)\}.\nonumber\end{align}
Note that if to put $\nu=\mu=3$ in Eqs.~\p{16}, \p{17}, one gets
the longitudinal spin $S^{l}$ and isospin $T^{l}$ correlation
functions, while setting $\mu,\nu=1,2$ gives the transverse spin
and isospin correlation functions \bal
S_{\mu\nu}^{t}(q)=-\frac{\delta_{\mu\nu}}{4}\bigl(I_f^{00}(q)+I_f^{03}(q)+
I_g^{30}(q)\bigr)\equiv\delta_{\mu\nu}S^{t}(q),&\nonumber\\
\mu,\nu=1,2,&\label{18}\\
T_{\mu\nu}^{t}(q)=-\frac{\delta_{\mu\nu}}{4}\bigl(I_f^{00}(q)-I_f^{03}(q)+
I_g^{30}(q)\bigr)\equiv\delta_{\mu\nu}T^{t}(q).&\nonumber\end{align}
The following relationships between the correlation functions hold
true \beqe S^{l}(q)=\frac{D(q)}{4},\quad
S^{{t}}(q)=T^{l}(q).\label{18.1}\end{equation} At zero temperature
and zero momentum transfer, the correlation functions satisfy the
sum rule \bal &S^{t}(q=0)=T^{l}(q=0)
\\&=-\frac{1}{2\pi^2\varrho}\int
dk\,k^2(f_{00}^2({k})+f_{03}^2({k})+g_{30}^2({k}))=
-\frac{1}{4}\,,\nonumber\end{align} where the r.h.s. is
independent of density and isospin asymmetry. Besides, the
transverse isospin correlation function satisfies the relationship
\bal T^{t}(q=0)&=-\frac{1}{2\pi^2\varrho}\int
dkk^2(f_{00}^2({k})-f_{03}^2({k})+g_{30}^2({k}))\nonumber\\ &=
-\frac{1-\alpha}{4},\end{align} where the r.h.s. is independent of
density.

\tbf{Correlation functions in nuclear matter with a $np$
condensate.} Further for numerical calculations we shall use the
effective zero range force, developed in Ref.~\cite{GSM} to
reproduce the pairing gap in $S=1,T=0$ pairing channel with Paris
NN potential:
\begin{equation} V({\bf r}_1,{\bf r}_2)=v_0\Bigl\{1-\eta\biggl(\frac{
\varrho(\frac{{\bf r}_1+{\bf
r}_2}{2})}{\varrho_0}\biggr)^\gamma\Bigr\}\delta({\bf r}_1-{\bf
r}_2)\label{9},\end{equation} where
$\varrho_0=0.16\,\mbox{fm}^{-3}$ is the nuclear saturation
density,  $v_0=-530\,\mbox{MeV}\cdot
\mbox{fm}^3,\,\eta=0,\,m=m_G$, $m_G$ being the effective mass,
corresponding to the Gogny force D1S. Besides, in the gap
equation~\p{8}, Eq.~\p{9} must be supplemented with a cut-off
parameter, $\varepsilon_c=60\,\mbox{MeV}$.   Just this set of
parameters, among total three parametrizations, used in
Ref.~\cite{GSM}, corresponds to the formation of bound states at
nonzero energy in low-density region of nuclear matter.

  \indent To find the correlation functions  one
should first solve the gap equation~\p{8} self-consistently with
Eqs.~\p{10}, \p{11}.  Then the correlation functions can be
determined directly from Eqs.~\p{12.3}, \p{16} and \p{17}. The
results of numerical determination of the energy gap as a function
of density for different asymmetries at zero temperature are shown
in Fig.~1. As one can see, with increasing asymmetry the magnitude
of the energy gap is decreased and the density interval, where a
$np$ condensate exists, shrinks to lower density. In reality
solutions exist for any $\alpha<1$ (the phase curves for larger
values of $\alpha$ are not shown in Fig.~1) and correspond to the
formation of BEC of deuterons at very low densities of nuclear
matter.
\begin{figure}[tb]
\includegraphics[height=7cm,width=7.5cm,trim=49mm 145mm 60mm 60mm,
draft=false,clip]{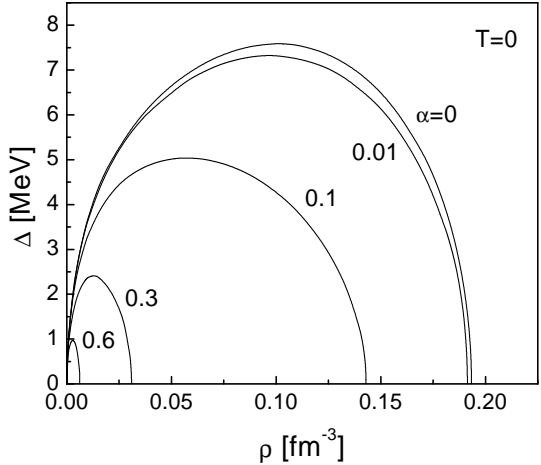} \caption{Energy gap as a function
of density at zero temperature and different asymmetries.
}\label{fig1}
\end{figure}

Now we consider the correlation functions $D({q})$ and
$S^{t}({q})$ for symmetric nuclear matter at zero temperature,
depicted in Fig.~2 (at $\alpha=0$, $T^{t}({q})=S^{t}({q})$). The
density correlation function changes sign at low momentum transfer
when the system smoothly evolves from the BEC regime to the BCS
one. These two regimes are distinguished by  negative and positive
values of the chemical potential $\mu$, respectively. In view of
Eq.~\p{18.1}, the longitudinal spin correlation function
$S^{l}({q})$  changes  sign through the BEC-BCS crossover as well.
The transverse spin correlation function, and, according to
Eq.~\p{18.1}, the longitudinal  and transverse isospin correlation
functions
 change fluently between BEC and BCS limits. The behavior of the density
 correlation function
 in isospin symmetric case at zero temperature qualitatively agrees with the behavior
 of the density correlation function in an ultracold fermionic atom gas with singlet
 pairing of fermions~\cite{MGB}. In Ref.~\cite{MGB},  the change in the sign of the
 density correlation function at low momentum transfer was considered as a signature of the
 BEC-BCS  crossover. We would like to extend their calculations
 taking into
 account the finite isospin asymmetry and finite temperature.
\begin{figure}[tb]
\includegraphics[height=7.7cm,width=8.2cm,trim=35mm 171mm 80mm 24mm,
draft=false,clip]{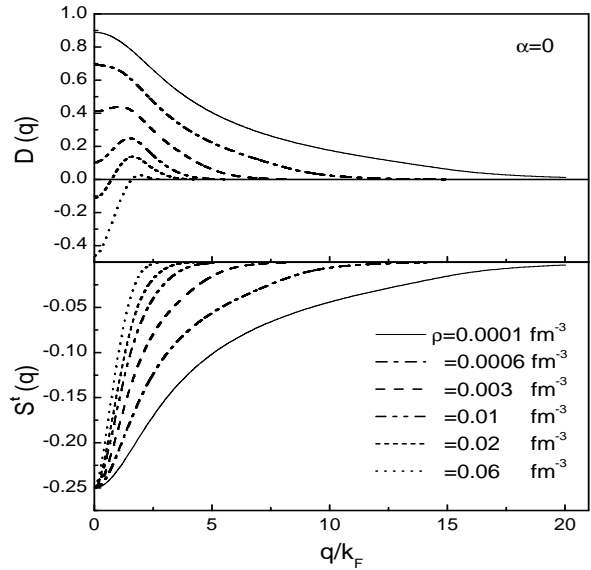} \caption{Density and transverse spin
correlation functions as functions of momentum at zero temperature
 and different densities for symmetric nuclear
matter.}\label{fig2}
\end{figure}

Fig.~3 shows the dependence of the density correlation function
$D({q}=0)$ at zero momentum transfer as a function of density for
a set of various isospin asymmetry parameters and zero
temperature. It is seen that with increasing the asymmetry
parameter the density correlation function  decreases. For strong
enough asymmetry, the function $D({q}=0)$ is always negative. In
accordance with the above criterion, the density region, where the
function $D({q}=0)$ has  positive or negative values, would
correspond to the BEC or BCS regime, respectively. Hence, as
follows from Fig.~3, for strong isospin asymmetry we would have
only the BCS state  for all densities where a $np$ condensate
exists. Obviously, this conclusion contradicts with the behavior
of the mean chemical potential $\mu$, being negative at very low
densities for any $\alpha<1$, and, hence, giving evidence to the
formation of BEC of bound states~\cite{IYB}. Thus, at strong
enough isospin asymmetry the criterion of the crossover, based on
the change of the sign of the density correlation function, fails
to  predict the transition to the BEC of deuterons in low-density
nuclear matter.

\begin{figure}[tb]
\includegraphics[height=6cm,width=8.2cm,trim=49mm 159mm 69mm 68mm,
draft=false,clip]{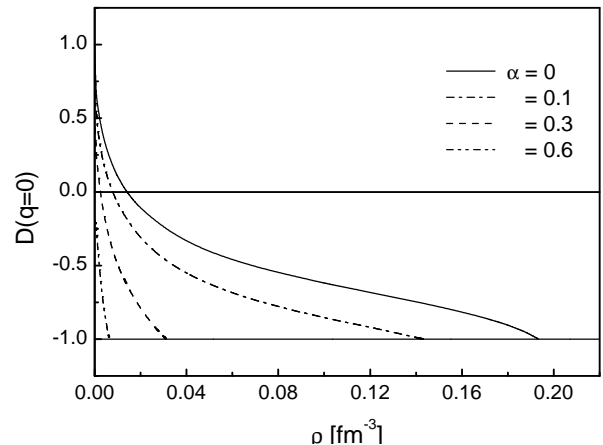} \caption{Density correlation function
  $D({\bf q}=0)$ as a function of density
at zero temperature for different isospin asymmetry parameters.}
\label{fig3}
\end{figure}

Now we consider  symmetric nuclear matter at finite temperature.
Fig.~4 shows the dependence of the density correlation function
$D( q=0)$ at zero momentum transfer as a function of density for a
set of various temperatures. It is seen that for not too high
temperatures the density response function is nonmonotonic and
twice changes  sign in the region of low densities. Hence, in
accordance with the above criterion,  we would have the density
interval $\varrho_1<\varrho<\varrho_2$ with the BEC state,
surrounded by the density regions with the BCS state. However,
this conclusion contradicts with the behavior of the mean chemical
potential $\mu$ for these temperatures, being a monotone function
of density and  indicating  the formation of a BEC state at low
densities ($\mu<0$) and a BCS state at larger densities ($\mu>0$).
Thus, at finite temperature the criterion of the crossover,
formulated in Ref.~\cite{MGB}, fails to provide the correct
description of the transition between two regimes.

\begin{figure}[tb]
\includegraphics[height=6cm,width=8.2cm,trim=49mm 159mm 67mm 68mm,
draft=false,clip]{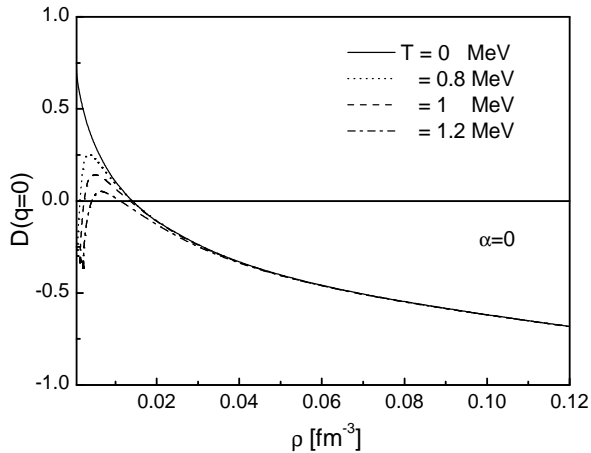} \caption{Density correlation function
  $D({\bf q}=0)$ as a function of density
at different temperatures for symmetric nuclear
matter.}\label{fig4}
\end{figure}
In summary, we have calculated the density, spin and isospin
correlation functions in superfluid nuclear matter with $np$
pairing correlations, intending to find the possible signatures of
the BEC-BCS crossover. It is shown that the transverse spin, and
longitudinal and transverse isospin correlation functions satisfy
the sum rule at zero momentum transfer and zero temperature, and
change smoothly between BEC and BCS regimes. In Ref.~\cite{MGB},
it was learned that the density correlation function in a
two-component ultracold fermionic atom gas with singlet pairing of
fermions changes sign at low momentum transfer across the BEC-BCS
transition, driven by a change in the scattering length of the
interaction at zero temperature. We have shown that for spin
triplet pairing the longitudinal spin correlation function plays
an analogous role to the density correlation function and changes
 sign at low momentum transfer across  the crossover in symmetric
nuclear matter at zero temperature. However, while giving a
satisfactory description of the density-driven BEC-BCS crossover
in dilute nuclear matter at zero temperature for the isospin
symmetric case, this criterion fails to provide the correct
description of the crossover at finite isospin asymmetry (nonequal
densities of fermions of different species) or finite temperature.
Hence, the criterion in Ref.~\cite{MGB} cannot be considered as
the universal indication of the BEC-BCS transition. During the
Mott transition, when the chemical potential changes sign, there
is a qualitative change in the quasiparticle energy spectrum: the
minimum shifts from a finite (BCS state) to zero-momentum value
(BEC state) (see Eq.~\p{6'''} and Ref.~\cite{PMT}). As such, the
presence (BCS) or absence (BEC) of the singularity in the momentum
distribution of the quasiparticle density of states represents the
universal signature of the BEC-BCS transition. This transition may
be relevant, and could give a valuable information on $np$ pairing
correlations, in low-density nuclear systems, such as  tails of
nuclear density distributions in exotic nuclei, produced at
radioactive nuclear beam facilities, expanding nuclear matter in
heavy ion collisions, low-density nuclear matter in outer regions
of neutron stars, etc.

\end{document}